\documentclass[preprint,apl]{revtex4-1}
\usepackage{graphicx}
\usepackage{dcolumn}
\usepackage{bm}
\usepackage{amsfonts,amssymb,amsmath,geometry, url,siunitx}
\usepackage{gensymb}
\usepackage{graphicx}
\usepackage{hyperref}
\usepackage{setspace}

\begin{document}
\title{Enhanced radial growth of Mg doped GaN nanorods: A combined experiment and first-principles based analysis} 

\author{Sanjay  Nayak}
\email{sanjaynayak@jncasr.ac.in}
\author{Rajendra Kumar}
\author{Nagaraja KK}
\affiliation{Chemistry and Physics of Materials Unit\\Jawaharlal Nehru Centre for Advanced Scientific Research (JNCASR), Bangalore-560064}
\author{S.M. Shivaprasad}
\email{smsprasad@jncasr.ac.in}
\affiliation{Chemistry and Physics of Materials Unit\\Jawaharlal Nehru Centre for Advanced Scientific Research (JNCASR), Bangalore-560064}
\date{\today}
\begin {abstract}
We discuss the  microstructural origin of enhanced radial growth in magnesium (Mg) doped gallium nitride (GaN) nanorods (NRs) using electron microscopy and \textit{first-principles} Density Functional Theory calculations. Experimentally, we find the  Mg incorporation increases surface coverage of the grown samples  and the height of NRs decreases as a consequence of an increase radial growth rate. We also observed the coalescence of NRs becomes prominent and the critical height of coalescence decreases with the increase in Mg concentration. From \textit{first-principles} calculations, we find the surface free energy of Mg doped surface reduces with increasing Mg concentration in the samples. The calculations further suggests a reduction in the diffusion barrier of Ga adatoms along [11$\overline{2}$0] on the side wall surface of the NRs, possibly the primary reason for the observed enhancement in the radial growth.
\end {abstract}


\maketitle

\section{Introduction}
Free standing semiconducting nanowires (NWs) and nanorods (NRs) are promising one-dimensional nanostructures having applications in various areas such as nanoelectronics, nanophotonics, nanosensing, etc\cite{tian2007coaxial,qian2004gallium,cui2001nanowire}. The fundamental, as well as the advanced studies on the growth and physical properties of such nanostructures, are of great interest within the scientific community\cite{persson2004solid,colombo2008ga}. The effective lateral stress relaxation in the case of nanostructures originating by the presence of facet edges can minimize or sometimes eliminate the formation of dislocations\cite{glas2006critical}.Typically, these nanostructures with radii in the order of several tens of nanometers and a length of micrometers are grown using modern epitaxial growth techniques like molecular beam epitaxy\cite{debnath2007mechanism} or metal-organic chemical vapor deposition\cite{kim2005zno}. 
\par
III-Nitrides being an important class of semiconductors have found in many optoelectronic and microelectronic devices available today. Due to the potential integration both the technologies, the growth of GaN on Si substrates attracts much attention. Also, GaN-on-Si is also a low-cost alternative to the conventional substrates like sapphire and SiC as the Si-based technology is comparatively mature and a successful integration gives extra advantages. In this direction, the epitaxial growth of III-V compound nanostructures including GaN on Si substrates has been extensively investigated\cite{ihn2006gaas,li2008growth,cerutti2006wurtzite}. To further increase the functionalities of the grown nanostructures, doping with suitable ions is partly necessary \cite{cui2001nanowire,wang2005ferromagnetism,tang2008vertically}. It is also observed in various cases; the process of doping may result in significant changes in the morphology as the presence of impurities on surfaces can affect the dynamics of adatoms arriving on it \cite{stamplecoskie2008general,radovanovic2007dopant,chen2009effect}. Therefore, the understanding of the microstructural origin of any change in the morphology owing to the incorporation of dopants is crucial for their use in various device applications.  GaN is an inherently n-doped compound semiconductor, and p-doping is a bit difficult. Out of many possible dopants, Mg is promising. Also, the Mg incorporation was also found to influence the morphology of the grown NRs strongly. A small amount of Mg can increase the tendency of the rods to coalesce while keeping their diameter unchanged without any broadening or tapering\cite{li2012gan}.
\par
Furtmayr \textit{et al.} studied the effect of Mg doping on MBE grown GaN NRs and showed an increase in Mg flux results in an increase in the diameter and a decrease in their height\cite{furtmayr2008nucleation}. The observed increase in the diameter may be because of the larger nucleation centers. However, it was concluded that the increase in the diameter of GaN NRs is due to the enhancement of the radial growth instead of larger nucleation sites. A similar increase in the radial growth in the case of Mg-GaN NRs with the increase in Mg incorporation has been reported by Zhang \textit{et al.}\cite{zhang2009influence} and Andrews \textit{et al.}\cite{wang2014p}. However, Bae \textit{et al.}\cite{bae2017selective} reported a variation of the height of NRs due to higher Mg flux instead of promoting vertical growth. Moreover, no convincing evidence for the origin of increased radial growth in MBE grown Mg-GaN NRs has been reported yet. Therefore, in this present work, we try to address this behavior by providing novel insights into the growth mechanism of Mg doped GaN NRs on Si by a combination of experimental and computational investigations.

\section{Methods}
\subsection{Experimental details}
The GaN NRs  were grown directly on Si (111) substrate under nitrogen rich conditions by a radio frequency plasma assisted molecular beam epitaxy (RF-PAMBE, SVTA-USA) system operating at a base pressure of $3\times 10^{-11}$ Torr. Prior to the growth, Si (111) substrate was ultrasonically cleaned in acetone for 10 minute. Degassing of the substrate at 600 $^o$C for 60 minute in the preparation chamber followed by degassing at 800 $^o$C for 30 minute and at 825 $^o$C for 5 minute in the growth chamber were carried out to get a clean Si ($7\times 7$) reconstruction. Before the growth, Si ($7\times 7$) reconstructed surface was exposed to metallic Gallium (Ga) for 10 sec. The temperature of the Ga effusion cell was maintained at 1060 $^o$C. A constant nitrogen flow rate of 4.5 sccm (standard cubic centimeter per minute), plasma forward power of 375 W, substrate temperature of 630 $^o$C and a growth duration of 4 hours were maintained for the growth of all samples. Mg fluxes were varied by adjusting the Mg K-cell temperature (see Table \ref{Mg flux}). The fluxes of the Mg and Ga were obtained from the beam equivalent pressure (BEP) and are tabulated in Table \ref{Mg flux}. Surface structural evolution was monitored \textit{in-situ} by reflection high energy electron diffraction (RHEED) and the morphology was determined \textit{ex-situ} using a field emission scanning electron microscope (FESEM, Quanta 3D operated at 20 kV). The crystal phase of the samples was determined using a high-resolution X-ray diffractometer (HR-XRD, Discover D8 Bruker) with a Cu K$_{\alpha}$ X-ray source with wavelength of 1.5406 \AA. \space Structure of the NRs at atomic scale was studied by high resolution transmission electron microscopy (HRTEM, FEI TITAN operated at 300 kV).  
\begin{table}[h]
\centering
\caption{Details of Mg flux rate}
\begin{tabular}{ c|c|c|c|c rrrrrrr}
 \hline
 Sample & Mg K-cell  & BEP  & Flux  & Mg:Ga\\
 Name & temp ($^o$C) &  (Torr) & (atoms cm$^{-2}$s$^{-1}$) & \\
 \hline 
 A & - & - & - & 0 \\  
 B & 340 & $4.5\times 10^{-9}$ & $1.29\times 10^{12}$ & 0.00818 \\
 C & 350 & $6.2\times 10^{-9}$ & $1.77\times 10^{12}$ & 0.01127 \\
 D & 360 & $8.9\times 10^{-9}$ & $2.52\times 10^{12}$ & 0.01618 \\
 \hline
 \hline   
\end{tabular}
\label{Mg flux}
\end{table}
\subsection{Simulation details}
\textit{Ab-initio} Density Functional Theory (DFT) simulations were carried to estimate the
surface energy and diffusion barrier of adatoms on side wall  surfaces of GaN NRs. A Generalized Gradient Approximation (GGA) of Perdew \textit{et al.}\cite{perdew1996generalized} was used for the exchange and correlation energy functional. Integrations over the Brillouin Zone of bulk w-GaN were sampled on a $\Gamma$-centered 5$\times$5$\times$3 uniform mesh of k-points in the unit cell of reciprocal space\cite{monkhorst1976special}. The lattice parameters and atomic co-ordinates were optimized  by using conjugate gradient algorithm to minimize the energy until the forces on each atom was less than 0.04 eV/\AA. Optimized lattice parameters of the unit cell of GaN are $a$ = 3.25 \AA \space and $c$ = 5.23 \AA.  The surfaces (both undoped and doped) were constructed within the slab model, where 12 atomic layers in [10$\bar{1}$0] direction were considered. With this model four layers of atoms in the middle were kept fixed at their bulk atomic positions to mimic the bulk configuration. We have used a vacuum of  16 \AA \space  perpendicular to (10$\overline{1}$0) surface to keep the interaction between image configurations weak. In case of surface cells, a single k-point was used for the long cell direction, while for other directions, k-points were chosen to  match the k-point density of the respective directions used in the bulk calculations.  We have estimated surface energy of doped (10$\overline{1}$0) surface by using following formula:
\begin{equation}
\mathrm{E_{surf(Ga_xMg_{1-x}N)}=\frac{1}{2A}[E_{slab(Ga_xMg_{1-x}N)}-E_{bulk(Ga_xMg_{1-x}N)}}]
\end{equation}
where A, E$\mathrm{_{slab}}$, and E$\mathrm{_{bulk}}$ correspond to the surface area of the slab, the total energy of the slab, and the total energy of the bulk with same Mg concentration as slab. The diffusion barrier of the Ga adatom on undoped and doped surface were estimated from total energy calculation of  respective configurations. Total energy of different configurations were estimated using  SIESTA code\cite{soler2002siesta}.
\section{Results and discussion}
\subsection{Experimental Results}
\begin{figure}[!ht]
\centering
\includegraphics[width=8cm]{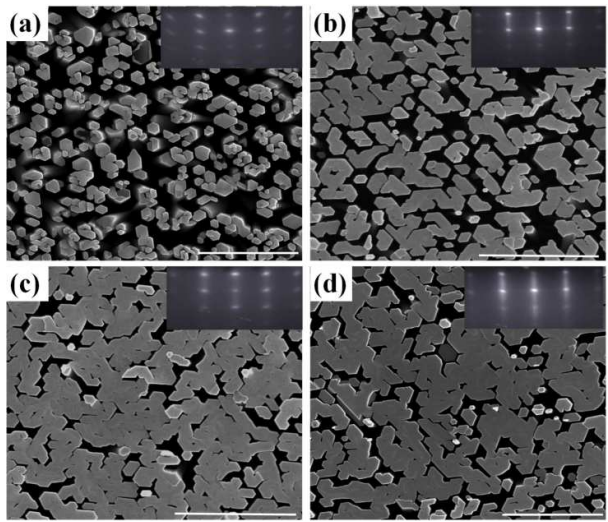}
\caption{(a)-(d) Plan view FESEM images of samples A, B, C and D, respectively. The scale bar is 2$\mu$m for all samples. Inset shows \textit{in-situ} RHEED pattern taken along $<11\overline{2}0>$  of  respective samples.}
\label{SEM}
\end{figure}
Plan view FESEM images of samples A-D grown with different Mg flux-rates (see Table \ref{Mg flux}) are shown in Fig.\ref{SEM}. We find that most of NRs are vertically aligned with the surface of substrate while few of them are slightly tilted for undoped sample. However, for doped samples (B, C and D), we observed well aligned NRs.  It is widely observed that, the relative misorientation of GaN NWs or NRs grown on Si (111) are typically about 3$^o$, both in-plane (twist) as well as out-of-plane (tilt)\cite{geelhaar2011properties,jenichen2011macro,wierzbicka2012influence,fernandez2014correlation}. The RHEED pattern shown as inset of the FESEM images confirm a typical of wurtzite structure and  spotty nature of it signifies a 3D morphology. However, we also observe the elongation of RHEED pattern along (10$\overline{1}$0) which may due to relative misorientation of NRs with each other. The RHEED pattern of doped samples (especially C and D) show an elongated spotty pattern with faint streaky lines which are attributed to  the electron scattering from the large flat c-plane tops of the NRs. A similar observation is also made by De \textit{et al.} \cite{de2016epitaxy}. To elucidate further, we estimate the surface coverage of samples A, B, C and D as 55$\pm$2\%, 65$\pm$2\%, 80$\pm$2\% and 80$\pm$2\% respectively  suggesting an increase in the surface coverage with increasing  Mg flux. 
\begin{figure}[!ht]
\centering
\includegraphics[width=8cm]{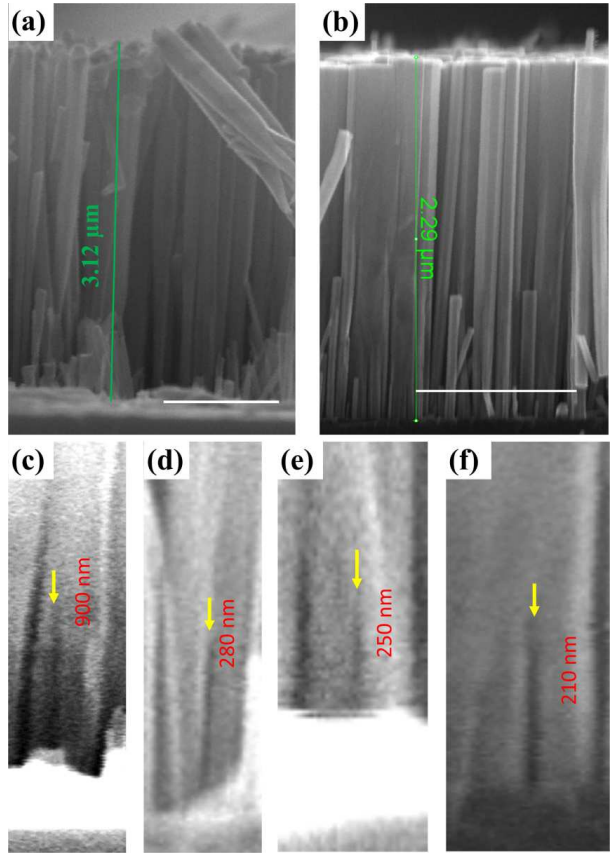}
\caption{(a) and (b) show cross section FESEM images of samples A and D. (c), (d), (e) and (f) represent the cross section FESEM image nearby GaN/Si interface of samples A, B, C and D respectively. Scale bar in (a) and (b) is 1$\mu$m.}
\label{SEM_cross_section}
\end{figure}
Further, to look at the interface of Si and GaN NRs, we carried out cross section FESEM imaging of  samples (see Fig.\ref{SEM_cross_section}). At the early stages of growth, NRs are quite isolated but they tends to coalesce as the growth proceeds. Interestingly, we find that with an increase in the Mg flux, the critical height $H_c$ for coalescence of NRs is decreasing as  $\approx$ 900$\pm$10 nm, 280$\pm$10 nm, 250$\pm$10 nm, 210$\pm$10 nm for sample A, B, C and D respectively. Kaganer \textit{et al.}\cite{kaganer2016nucleation} suggest H$_c$ of coalescence of NRs can be calculated by
\begin{equation}\label{kaganer}
H_c=4\left(\frac{E}{9\gamma\omega}\frac{I_1I_2}{I_1+I_2}l^2\right)^{1/4}
\end{equation}
where E is the Young modulus, $\gamma$ the surface energy, $\omega$ the width of the contact area (for the calculations, is equal to the radius of the thinner nanowire) and $I_i$ indicates the geometrical moments of inertia of the cross-section of the corresponding nanowire (i=1,2). For a cylinder, $I=\pi R^4$ where $l$ is separation between nanowires at the bottom. Equation \ref{kaganer} suggests that, while keeping all other parameters fixed an increase in $\gamma$ results in the reduction of $H_c$. Our first-principles calculations suggest a small reduction in the surface free energy (as shown in Fig.\ref{diffusion}(c)) of  (10$\overline{1}$0) with Mg incorporation, thus we infer in this case, the role of change in $\gamma$ is almost negligible in determining $H_c$. Further, it can also be inferred that an increase in the radius of NRs results into higher $H_c$, suggesting the mechanism proposed by Kaganer \textit{et al.}\cite{kaganer2016nucleation} may not be appropriate for the present case.
\par
\begin{figure}[!ht]
\centering
\includegraphics[width=8cm]{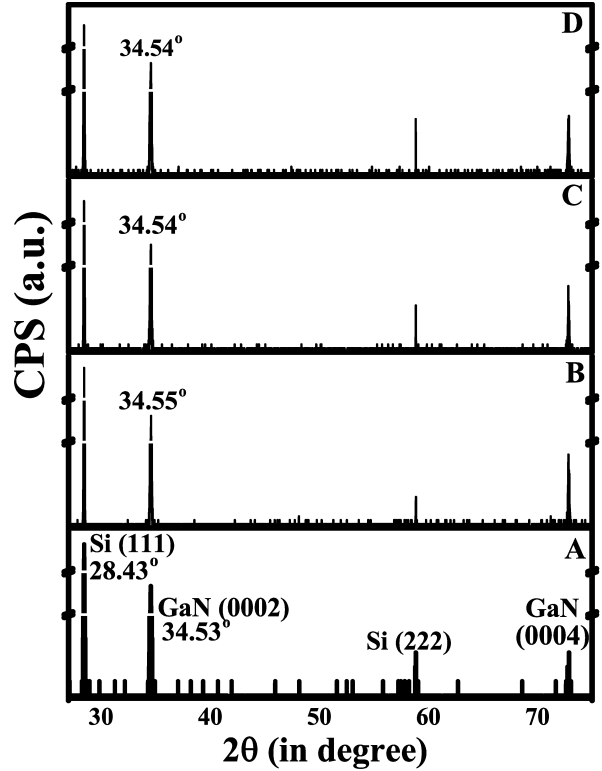}
\caption{Symmetric 2$\theta$-$\omega$ scan acquired by HRXRD of samples A-D, respectively. }
\label{XRD}
\end{figure}
To study the structural properties of Mg doped GaN NRs, we carried out HR-XRD measurements of  samples (see Fig~\ref{XRD}). From the symmetric 2$\theta-\omega$ scan, we find GaN (0002) reflex,  along with Si (111), (222) and GaN (0004) reflexes are present (see Fig.\ref{XRD}). Figure \ref{XRD} suggests all the GaN NRs samples (undoped and Mg-doped) possess single crystallinity, wurtzite structure with a preferential growth direction along the c-axis (0001) with the epitaxial relation of [111]$_{Si}$ $||$ [0002]$_{GaN}$. The intense GaN (0002) reflexes  at 34.54$^o$, 34.55$^o$, 34.54$^o$, and 34.54$^o$ for samples A, B, C and D respectively  suggest with Mg incorporation, c lattice parameter of GaN has reduced, contrary to the  typical behavior  wherein the lattice parameters of GaN increases as ionic radius of Mg is larger than that of Ga\cite{kirste2013compensation}. However, if defect such as complexes of N-vacancy are formed, the local strain will change \cite{kirste2013compensation}and results in the reduction of lattice parameters.
\begin{figure}
\centering
\includegraphics[width=12cm]{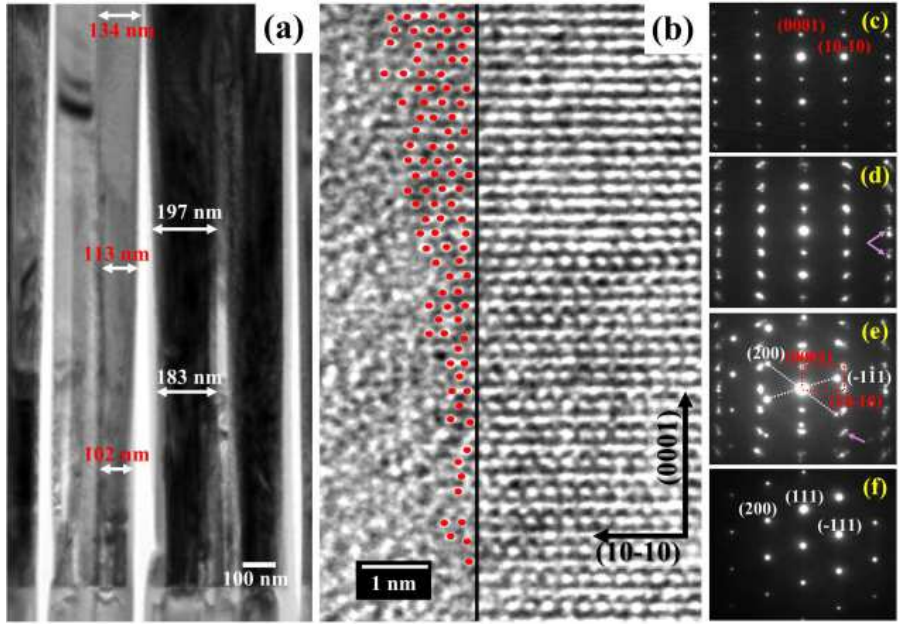}
\caption{(a) shows bright-field TEM image of sample B. (b) shows HRTEM image of a single NRs. (c), (d), (e) and (f) represent SAED pattern recorded from top, just above GaN/Si interface, GaN/Si interface and  substrate.}
\label{TEM}
\end{figure}
\par
To look at the atomistic origin of coalescence of NRs, we obtained HR-TEM image of samples B and are shown in Fig.\ref{TEM}. In Fig.\ref{TEM} (a) the dimensions of two different NRs are shown at different height  concluding tapering effect in the NRs sample due to the formation of atomic steps (see red dots in Fig.\ref{TEM}(b)). Figure \ref{TEM} (a) also shows high densities of NRs are formed.  Figure \ref{TEM}(c) shows SAED pattern at top part of the NRs where a pattern which is perfectly single crystalline in nature was recorded. However, as we move towards the interface, we started observing other diffraction spots including expected ones (see arrow marked spots in Fig.\ref{TEM} (d) and (e)). As these extra spots are lying on same circle with the other expected spots, we conclude that the extra pattern are due to mosaic nature of NRs. From the SAED pattern of the interface we found that the epitaxial relation between GaN NRs and Silicon substrate are [0002]$_{GaN}$ $||$ [111]$_{Si}$ and [1$\overline{2}$10]$_{GaN}$ $||$ [01$\overline{1}$]$_{Si}$. Absence of any ring pattern in the SAED pattern obtained from interface region further indicates that no amorphous region is formed at the GaN/Si interface.  
\par
The observation of high densities of NRs as well as the tapering effect  are the main reasons for the coalescence of NRs in this particular case. Along with these factors, the reduction in the values of H$_c$ with an increase in the Mg concentration together suggest that the radial growth rate of sidewall surface of NRs is increasing with an increase in the Mg incorporation. To find the atomistic origin of this enhanced radial growth of NRs we have carried out the first-principles calculations and the next section of the work is dedicated to same.
\subsection{Estimation of Diffusion Barrier}
\begin{figure}
\centering
\includegraphics[width=12cm]{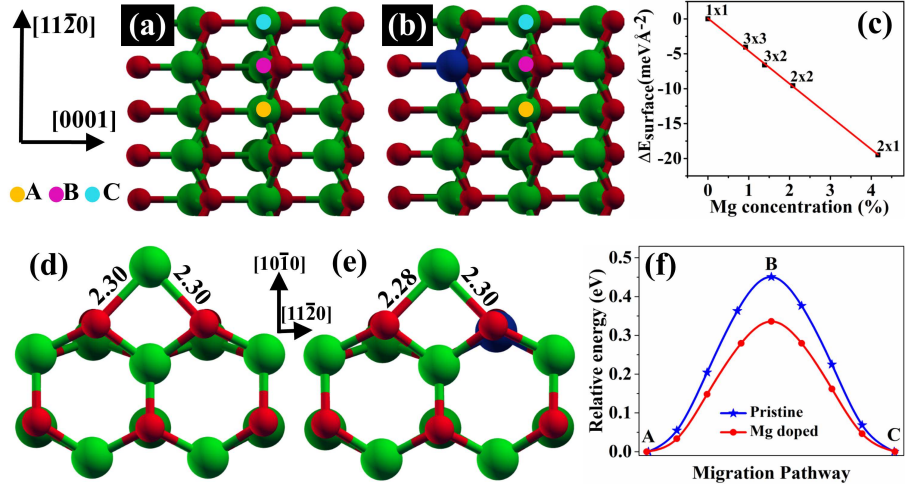}
\caption{(a) and (b) show the  undoped and Mg doped  slab  models  used  for calculations, respectively. (c) represent the estimated surface energy of the doped surface with different doping concentrations (the number mention nearby data points represent in-plane supercell dimension used for the particular calculation). (d) and (e) represent crossectional view of relaxed atomic structure of Ga adatom at undoped and doped surface at most stable configurations of respective cases. (f) represent the relative energy of different configurations where ad-atoms were kept at various position along the path A-B-C. The lowest energy configurations are the reference ones.}
\label{diffusion}
\end{figure}

 It is estimated that for (10$\overline{1}$0) (or `m') plane of GaN, which forms the side wall surfaces of NRs, the diffusion barrior of Ga-adatom shows an anisotropic value. The diffusion barrier along [11$\overline{2}$0] is merely 0.21 eV while along [0001] direction it is 0.93 eV, \cite{lymperakis2009large} which promotes the radial growth rate of NRs. To study the role of the presence of Mg atoms on the diffusion barrier of  adatoms, we estimated the total energy of  various configurations where Ga adatoms were kept manually at various position of (10$\overline{1}$0) surface as shown in Fig.\ref{diffusion}(a) and (b) wherein a 3$\times$2 in-plane supercell was used. We have taken the similar path in both cases (see Fig.\ref{diffusion}) for the estimation of diffusion barrier of Ga adatoms along [11$\overline{2}$0] as discussed in work of Lymperakis \textit{et al.}\cite{lymperakis2009large} and Jindal \textit{et al.}\cite{jindal2010computational}. Before we discuss about the calculation regarding the estimation of diffusion barrier of Ga adatoms on both the undoped and doped surface, we provide a brief discussion on the atomic structure of both the surfaces. In the relaxed geometry of undoped (10$\overline{1}$0) surface we observed that Ga atoms at top layer of the slab moves inward into the bulk whereas N atoms moves outward into vacuum  in comparison to the ideal (10$\overline{1}$0) cleaved surface, which results in the vertical separation of $\approx$ 0.43 \AA \space along $\langle10\overline{1}0\rangle$ and buckling of surface Ga-N bond by 13.45$\degree$. Because of a such relaxation, we find Ga-N bond reduced to 1.86 \AA \space which is contracted by $\approx$ 6 \% w.r.t. the same in bulk.  Such a behavior is quite consistent with the other reports in the literature\cite{landmann2015gan,gonzalez2014comparative}. In case of relaxed surface of Mg doped (10$\overline{1}$0) surface, the atomic relaxation of Ga, Mg and N is bit complex. While  Ga atoms moves inward to the bulk,  N atoms making bond only with Ga atoms move outward to vacuum. The Mg atoms displaced towards the bulk by 0.10 \AA \space and the N atom that makes bond with Mg atoms moves inward to the bulk by 0.03 \AA. The Mg-N bond length at (10$\overline{1}$0) surface is $\approx$ 1.98 \AA, while in bulk it is 2.05 \AA, which suggests Mg-N bond length shrinks at surface in comparison to bulk by $\approx$ 3.4\%.
\par
A minimum energy configuration is attained when Ga adatom takes ``A" site (see yellow dot in Fig.\ref{diffusion} (a)) where it make bond with two surface N atoms with a bond length of 2.30 \AA (see Fig.\ref{diffusion} (d)). From Mulliken charge analysis we find in this configuration the charge transfer from Ga adatom to the slab is $\approx$ 0.56$|e|$. Similarly, with Mg doped (10$\overline{1}$0) surface the minimum energy configuration is when Ga adatoms takes ``A" site. We find at this configuration the Ga-N bond length are 2.28 \AA \space and 2.30 \AA \space (see Fig.\ref{diffusion} (e)) and the charge transfer is $\approx$ 0.59$|e|$. To estimate the diffusion barrier, we calculated the total energy of various configurations where Ga adatom migrate from one minimum energy configuration (site ``A") to another (site ``C"). At undoped surface, we estimated the diffusion barrier as $\approx$ 0.45 eV, while with Mg doped it reduced to 0.33 eV. Such a reduction in the diffusion barrier will increase the diffusion length (L$_{diff}$) of adatoms as L$_{diff}$ is proportional to $\sim \sqrt{\Gamma \tau}$, where $\tau$  represent the average adatom life time.  $\Gamma$ is the diffusion coefficient which is proportional to $\sim exp(-E_{diff}/k_BT)$, where $E_{diff}$ is the diffusion barrier, $k_B$ and $T$ are the Boltzmann constant and absolute temperature. By taking the growth temperature  as 900 K, we find L$_{diff}$ of Ga adatoms on Mg doped surface (with surface Mg concentration as 8.3\%) is two times higher than the undoped (10$\overline{1}$0) surface of GaN. Such an increase in the L$_{diff}$ is due to the presence of Mg on surface which increases the probability of incorporation of adsorbed Ga adatoms leading to the enhanced radial growth of NRs.
 \par
Work of Kaganer \textit{et al.}\cite{kaganer2016nucleation} suggests that, the nucleation of nanowires are random and homogeneous. After the nucleation, such nanowires attain a self equilibrated diameter beyond  which nanowires  grow only in the axial direction but not in the radial direction. We propose such a behavior of NRs growth is the reason for the pinning of surface coverage at 80$\pm$2\% for higher Mg-doped samples.

\section{Summary}
In summary, we have synthesized and characterized  Mg doped GaN NRs grown on Si (111) surface. We have observed that with an increase in the Mg concentration the surface coverage of the samples increases due to higher growth rate of the sidewall surface of NRs. From TEM analysis we have found that the NRs are tapered as a consequence of the formation of atomic steps on the side surfaces of NRs. From SAED analysis we have found that the NRs  nearby interface are more mosaic in nature while  mosaicity decreases with the increase in the thickness. From first-principles calculations we have concluded that reduction in the diffusion barrier of Ga ad-atoms due to presence of Mg atoms, on [10$\overline{1}$0] surface, along [11$\overline{2}$0] direction  is the primary reason for the higher radial growth rate of NRs.
\bibliographystyle{apsrev4-1}
\bibliography{Mg-Si}
\end{document}